\documentclass[
showpacs,
floatfix,
aps,
prb,
amsmath,
nofootinbib,
twocolumn,
superscriptaddress,
]{revtex4}

\usepackage{graphicx}
\usepackage{dcolumn}
\usepackage{amsmath}
\usepackage[varg]{txfonts}

\def\oc{\omega_{\mbox{\tiny {C}}}}
\def\rc{R_{\mbox{\tiny {C}}}}

\begin{document}
\title{\bf Magnetotransport in a two-dimensional electron system in dc electric fields}
\author{W. Zhang}
\author{H.-S. Chiang}
\affiliation{School of Physics and Astronomy, University of Minnesota, Minneapolis, Minnesota 55455, USA} 
\author{M. A. Zudov}
\email[Corresponding author. Email: ]{zudov@physics.umn.edu}
\affiliation{School of Physics and Astronomy, University of Minnesota, Minneapolis, Minnesota 55455, USA} 
\author{L. N. Pfeiffer}
\author{K. W. West}
\affiliation{Bell Laboratories, Lucent Technologies, Murray Hill, New Jersey 07974, USA}
\received{8 August 2006; revised manuscript received 8 November 2006}

\begin{abstract}
We report on nonequilibrium transport measurements in a high-mobility two-dimensional electron system subject to weak magnetic field and dc excitation.
Detailed study of dc-induced magneto-oscillations, first observed by Yang {\em et al}., reveals a resonant condition that is qualitatively different from that reported earlier.
In addition, we observe dramatic reduction of resistance induced by a weak dc field in the regime of separated Landau levels.
These results demonstrate similarity of transport phenomena in dc-driven and microwave-driven systems and have important implications for ongoing experimental search for predicted quenching of microwave-induced zero-resistance states by a dc current.

\end{abstract}
\maketitle

Nonequilibrium magnetotransport in very high Landau levels (LLs) of two-dimensional electron systems (2DESs) is of intense current interest.
Major efforts, both theoretical and experimental, have been directed toward microwave photoresistance phenomena, such as microwave-induced resistance oscillations\citep{zudov:201311,ye:2193} (MIRO) and zero-resistance states\citep{mani:646,zudov:046807} (ZRS).
Conversely, other novel effects observed in 2DESs, not irradiated by microwaves, have not received due attention. 
These include magneto-oscillations from interface acoustic phonon scattering\citep{zudov:3614} and those from Zener tunneling between tilted LLs,\citep{yang:076801} both relying on large-angle scattering required by momentum and/or energy conservation.

Experimentally, MIRO appear in photoresistance, $\Delta R(\omega, B) \equiv R(\omega,B)-R(0,B) \propto \sin(2\pi\omega/\omega_c)$, where $\omega=2\pi f$ is the microwave frequency, $\oc=eB/m^*$ is the cyclotron frequency of an electron, and $B$ is the magnetic field.  
MIRO were initially explained in terms of impurity scattering,\citep{durst:086803,vavilov:035303,ryzhii:2078} but it is currently believed that the electron distribution function effects play a dominant role.\citep{dmitriev:115316}
The fact that MIRO minima evolve into ZRS is linked to microscopic negative resistivity and its instability, which results in formation of current domains.\citep{andreev:056803}
While the concept of negative resistivity has recently found some support in bichromatic experiments,\citep{zudov:236804} the conjecture that ZRS only exist below some critical current density\citep{andreev:056803} has not been experimentally verified.
To systematically approach this problem, it is important first to better understand the effects of the dc current on 2DESs without microwaves, as it itself can strongly modify magnetotransport properties.\citep{yang:076801}


In this paper we report on magnetotransport measurements in a high-mobility 2DES under dc current excitation driving the system into a nonequilibrium state. 
Our sample was cleaved from a symmetrically doped GaAs/Al$_{0.24}$Ga$_{0.76}$As 300-\AA-wide quantum well grown by molecular beam epitaxy.
A Hall bar mesa of a width $w=100$ $\mu$m was fabricated using photolithography. 
Ohmic contacts were made by evaporating Au/Ge/Ni and thermal annealing in forming gas ambient.
The experiment was performed in a $^3$He cryostat, equipped with a superconducting solenoid, at a constant coolant temperature $T\simeq 1.5$ K.
After illumination with visible light, electron mobility $\mu$ and density $n_e$ were $\simeq$\,$1.2 \times 10^7$ cm$^2$/Vs and $3.7 \times 10^{11}$ cm$^{-2}$, respectively.
The differential resistance $r(I,B)\equiv dV/dI$ was measured using low-frequency lock-in amplification both in changing $B$ at fixed dc current $I$ and in changing $I$ at fixed $B$.

By analogy with microwave-driven 2DESs, we express the resistance under dc excitation as $R(I,B)\!\equiv\!R(0,B)\!+\!\Delta R(I,B)$, where $R(0,B)$ is the linear-response resistance and $\Delta R(I,B)$ is the dc-induced, nonlinear change.
The latter translates to $\Delta r(I,B) = \partial _I[I\Delta R(I,B)]_B$. 
In the original work by Yang {\it et al.}\citep{yang:076801} it was found that $\Delta r$ acquires oscillations as a function of $\epsilon\,=\,2\rc/\Delta Y$, owing to the commensurability of two length scales\,$-$\,the cyclotron diameter $2R_c\!=\!2 v_F/\omega_c$\,($v_F$ is the Fermi velocity) and the inter-LL spacing $\Delta Y=\hbar \oc /eE$ ($E=IB/en_ew$ is the Hall field).
After mapping these length scales onto energy scales, one obtains $\epsilon = \omega_H/\oc$, where $\hbar \omega_H= \gamma\hbar (2\pi/n_e)^{1/2}I/ew$ is the energy associated with the Hall voltage drop across the cyclotron orbit and $\gamma \simeq 2$ is the fitting parameter.
Notice that for a given sample $\omega_H$ is controlled by $I$ only.
Oscillations were explained in terms of large-angle elastic scattering between Hall field-tilted LLs,\citep{yang:076801} and, for brevity, will be referred to as Hall field-induced resistance oscillations (HIRO).
Recently, HIRO were observed in independent experiment.\citep{bykov:245307}

Experiments in dc-driven 2DESs are unique, and thus important, in several respects.
First, one can easily cover a wide ``frequency'' range (in this study, $0\leq \omega_H/2\pi \leq 120$ GHz) and perform controlled ``frequency'' sweeps, which are difficult in microwave experiments.
Second, one can easily explore the regime of $\epsilon\!\rightarrow\!0$, which normally would require $B\!\rightarrow\!\infty$.
Finally, a dc-driven 2DES appears intrinsically simpler (e.g., intensity, reflection, etc. are absent), so one expects more transparent comparison with theory.

\begin{figure}[t]
\includegraphics{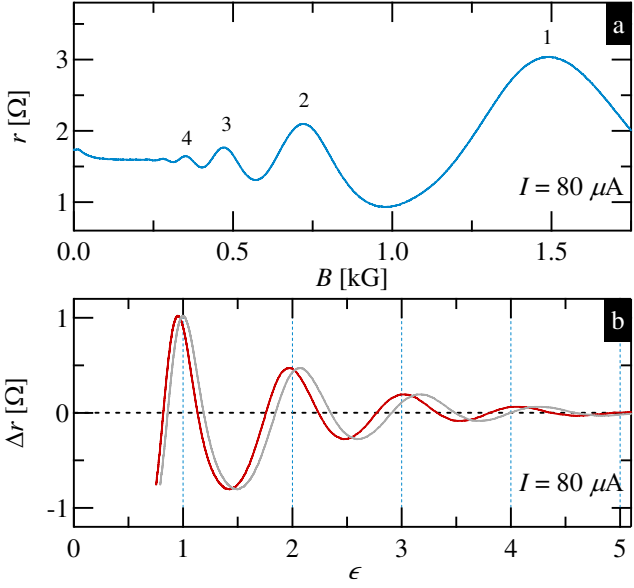}
\caption{(color online) 
(a) Differential resistance $r(B)$ at $I\!=\!80$ $\mu$A.
(b) Correction to differential resistance $\Delta r (\epsilon)$ extracted from (a) using $\gamma =1.9$\, [red (dark)] and $\gamma =2.0$\, [gray (light)].
}
\label{b}
\end{figure}

In Fig.\,\ref{b}(a) we present differential magnetoresistance $r$ measured at $I=80$ $\mu$A ($\omega_H/2\pi \simeq 65$ GHz) which shows HIRO up to fifth order.
From these data we extract the dc-induced correction $\Delta r$
which is plotted in Fig.\,\ref{b}(b) as a function of $\epsilon$.
According to the method developed for analysis of MIRO waveforms,\citep{zudov:041304} the period is determined from the higher-$\epsilon$ oscillations and the phase is obtained by extrapolation to $\epsilon=0$.
Using $m^*=0.0635 m_0$ obtained in a separate MIRO experiment, this procedure yields $\gamma = 1.9$, which falls in the middle of a rather wide range (1.63 to 2.18) of $\gamma$ reported earlier.\citep{yang:076801,bykov:245307} 
The result is plotted as a dark curve showing that $\Delta r$ maxima\,(minima) occur close to integer\,(half-integer) $\epsilon$.
This is qualitatively different from previous reports\citep{yang:076801,bykov:245307} which ascribed maxima in {\em total} resistance to integer $\epsilon$.\citep{note2}
For later comparison with theory, the gray curve is plotted in Fig.\,\ref{b}(b) using $\gamma=2.0$; while the first peak appears at $\epsilon \simeq 1$, the phase shift accumulates with $\epsilon$.

The difference between the resonance condition obtained here and that in the earlier reports is likely to reflect different sample quality and procedures used to analyze the data. 
While both methods can reasonably describe low-order oscillations, it was previously shown\citep{zudov:041304} for MIRO that extracting the period and phase from the higher-order oscillations is more reliable. 
These oscillations, however, may remain unresolved in lower-mobility samples, such as those used in early experiments.
We also note that the measured value of the effective mass is a few percent lower than the usually quoted value of $0.067m_0$.\citep{sze}
Although similar mass reduction was recently observed in other laboratories using different samples and setups,\citep{zudov:041304,mani:146801} its origin remains unclear and presents an interesting problem for future studies.

\begin{figure}[t]
\includegraphics{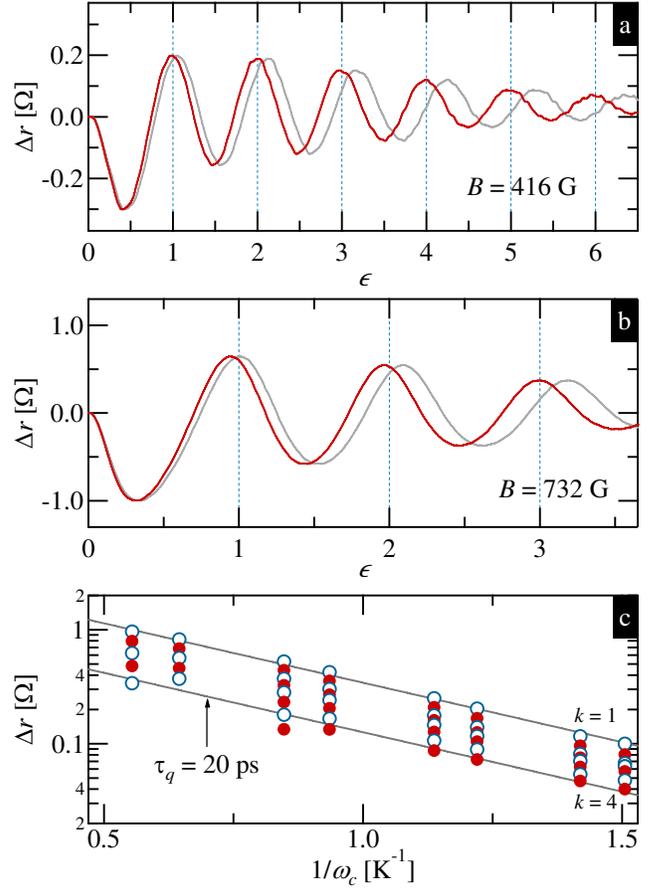}
\caption
{(color online)
Correction to differential resistance $\Delta r(\epsilon)$ obtained in $I$ sweeps at fixed $B=416$ (a) and $732$ G (b) using $\gamma=1.9$\, [dark (red)] and $\gamma=2.0$\, [gray (light)].
(c) $\Delta r(1/\oc)$ at the first four\,($k=1,2,3,4$) maxima\, [solid (red)] and minima\, [open (blue)]. 
Lines correspond to $\tau_q \simeq$ 20 ps, consistent with the decay in Fig.\,\ref{b}.
}
\label{i}
\end{figure}

We now turn to the discussion of the data recorded at fixed $B$ while sweeping $I$\citep{yang:thesis}.
In Figs.\,\ref{i}(a) and 2(b) we plot $\Delta r(\epsilon)$, obtained in $I$ sweeps at $B=416$ and $732$\, G, respectively, for $\gamma = 1.9$\, [dark (red)] and $\gamma = 2.0$\, (gray).
For $\gamma=1.9$\citep{note3} we again observe that apart from small deviations at $\epsilon \lesssim 1$, the positions of $\Delta r(\epsilon)$ maxima\,(minima) are well described by integer\,(half-integer) $\epsilon$.
The observed slower decay of HIRO is presumably explained by a constant Dingle factor $\delta = \exp(-\pi/\oc \tau_q)$, which dominates the decay in $B$ sweeps such as that shown in Fig.\,\ref{b}(b).
As called for by further analysis, we estimate quantum scattering time $\tau_q$ from the Dingle plots shown in Fig.\,\ref{i}(c) for the first four maxima\, [solid (red) circles] and minima\, [open (blue)].
Solid lines corresponding to $\tau_q \simeq$ 20 ps describe the data reasonably well. This value is also consistent with the decay observed in Fig.\,\ref{b}(b).
\begin{figure}[t]
\includegraphics{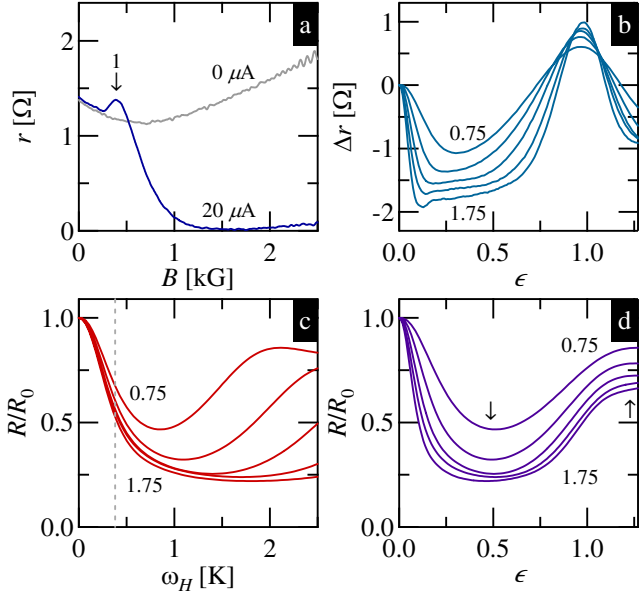}
\caption
{(color online)
(a) Differential magnetoresistance $r(B)$ at $I=20$ $\mu$A.
(b) Correction to differential resistance $\Delta r (\epsilon)$ obtained from $I$ sweeps at fixed $B$ (0.75 to 1.75 kG in 0.25 kG steps).
(c) Normalized total resistance $\Delta R(\omega_H)/R_0$ reconstructed from data in (b). 
Vertical line marks LL width $2\Gamma=\hbar/\tau_q = 380$ mK.
(d) Vertical arrows mark the first minimum\, (maximum) at $\epsilon \simeq 1/2$\, ($\epsilon \simeq 5/4$) in $\Delta R(\epsilon)/R_0$.
}
\label{sll}
\end{figure}

\begin{figure}[t]
\includegraphics{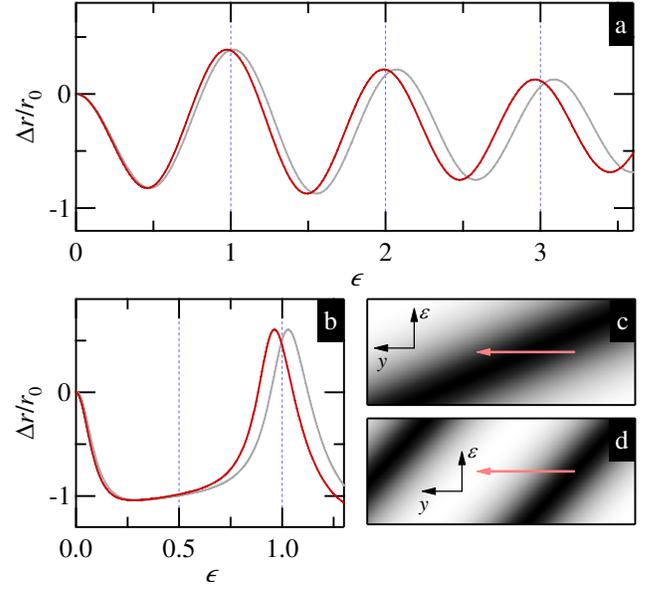}
\caption{(color online)\,(a)\,gray curve is calculated correction to differential resistance for $\oc \tau_q=4$ and $\gamma=2$. 
Dark (red) curve is obtained by compressing $\epsilon$ axis by 5\% to simulate $\gamma=1.9$.
(b)\,Same as (a), but for $\oc \tau_q=10$. 
(c) and (d) show density of states (DOS) (Landau levels) at $\omega_H \simeq \Gamma/\hbar$ and $\omega_H \simeq \oc/2$, respectively.
Horizontal arrows (length is equal to cyclotron diameter) represent characteristic electron transitions. 
}
\label{fig4}
\end{figure}

Now we turn our attention to a further experimental regime in dc-driven 2DESs, i.e., $\omega_H <\oc$ and $2\Gamma<\hbar\oc$, which is particularly relevant for the ongoing search for the domain current associated with microwave-induced ZRS.
In Fig.\,\ref{sll}(a) we present differential magnetoresistance under relatively weak dc excitation, $I=20$ $\mu$A ($\omega_H/2\pi \simeq 16$ GHz) plotted together with the zero-bias trace.
While only one HIRO peak is resolved (see\,$\downarrow$), dramatic reduction in $r$ is observed at $\oc > \omega_H$.
One immediately notices a striking similarity to the microwave-induced suppression of resistance recently observed at $\oc\!> \!\omega$,\citep{willett:026804,dorozhkin:201306,zudov:041303} which was related to intra-LL absorption. 

dc-induced suppression is best studied in $I$ sweeps which allow easy access to small $\epsilon$.
The data obtained at selected $B$ (from 0.75 to 1.75\,kG, in 0.25\,kG steps) are plotted in Fig.\,\ref{sll}(b).
As $B$ increases, the first minimum in $\Delta r$ becomes deeper and shifts to smaller $\epsilon$.
Figure 3(c) shows that total resistance, plotted as a function of $\omega_H$, is suppressed by about a factor of 5.
One also observes that the width of the zero-bias peak becomes smaller and seems to saturate at higher $B$.
These observations can be linked to growing suppression of scattering within one LL, as its real-space width $2\Gamma/eE$ approaches and eventually becomes smaller than $2\rc$ [cf.\,Fig.\,\ref{fig4}(c)].
The crossover point $\hbar \omega_H\!=\!2\Gamma\!=\!\hbar/\tau_q$ is marked by a vertical dashed line in Fig.\,\ref{sll}(c), in good agreement with the half-width of the zero-bias peak.
With proper theoretical input, this might provide a convenient way to extract $\tau_q$.

We now convert the horizontal axis to $\epsilon$ and replot the data in Fig.\,\ref{sll}(d).
While the minimum becomes wider and deeper with increasing $B$, i.e., with increasing separation of LLs, it remains roughly centered at $\epsilon \simeq 1/2$\, (see $\downarrow$).
This marks the middle of the region $2\Gamma \lesssim \omega_H \lesssim \oc-2\Gamma$, where both intra- and inter-LL transitions are strongly suppressed [cf.\,Fig.\,\ref{fig4}(d)].
At higher $\omega_H$, inter-LL scattering increases and the resistance starts to grow peaking at $\epsilon \simeq 5/4$\, (see $\uparrow$). 

To qualitatively explain our observations we propose a simple model based on impurity scattering, used to explain MIRO,\citep{durst:086803} modified for strong dc-excitation. 
The local current density $J_y$, which enters the resistivity $\rho_{xx}\!\simeq\!\rho_{xy}^2\sigma_{yy}\!\simeq\!\rho_{xy}^2J_y/E$, is given by
\begin{align}
J_{y}(\textbf{r})=-2&\pi e\int d\varepsilon \ d^2\Delta \textbf{r} [f(\textbf{r},\varepsilon)-f(\textbf{r}',\varepsilon)]\notag \\ 
&\times \nu(\textbf{r},\varepsilon)\nu(\textbf{r}',\varepsilon) M(\textbf{r},\textbf{r}')\Delta y.
\label{t1}
\end{align}
Here, $\varepsilon$ is the energy of an electron, $\textbf{r}$\,($\textbf{r}'$) is its initial\,(final) position, $\Delta \textbf{r}\!\equiv\!\textbf{r}'\!-\!\textbf{r}\!\equiv\!(\Delta x,\,\Delta y)$, $f(\textbf{r},\varepsilon)$ is the distribution function, $\nu(\textbf{r},\varepsilon)$ is the local density of states\,(DOS), and $M(\textbf{r}',\textbf{r})$ is the scattering rate with the DOS factored out.  
The dc current along $x$ direction induces Hall electric field $E$ which tilts the LLs in the $y$ direction: $\varepsilon\!\simeq\!\hbar \oc (N+1/2)\!-e\!Ey$ and therefore $f(\textbf{r}',\varepsilon)\!=\!f(\textbf{r},\varepsilon\!-\!eE\Delta y)$, $\nu(\textbf{r}',\varepsilon)\!=\!\nu(\textbf{r},\varepsilon\!-\!eE\Delta y)$.
For simplicity, we assume constant $M$, limit hopping distance to $\leq 2\rc$, and consider overlapping LLs, $\nu(\varepsilon)\!=\!(m^*/2\pi\hbar^2)[1\!-\!2\delta \cos(2\pi \varepsilon/\hbar\oc)]$, with $\delta\!=\!\exp (-\pi /\oc\tau_q)$.
At $k_BT\!\gtrsim\!eE(2\rc)$, $f(\varepsilon)$ is a slow function and $f(\varepsilon)-f(\varepsilon-eE\Delta y)\simeq-\partial_\varepsilon f(\varepsilon)eE \Delta y$.  
After the integration of Eq.\,(\ref{t1}), correction to differential resistivity is found to be
\vspace{-0.05in}
\begin{equation}
\frac{\Delta r}{r_{D}}=-\frac{4\delta^2}{\pi^2\epsilon^2}\left[(4\pi^2\epsilon^2-9)J_{2}(2\pi \epsilon)+4\pi\epsilon J_1(2\pi\epsilon)\right],
\label{dr}
\end{equation}
where $\rho_D$ is the Drude resistivity at $B\!=\!0$ and $J_k(x)$ is the Bessel function.

In Fig.\,\ref{fig4}(a), we plot the relative correction to the differential resistance, extracted from Eq.\,(\ref{dr}), defined as $(r_{xx}-r_0)/r_0$, where $r_0=r_D(1+2\delta^2)$, for $\tau_q \oc =4$ using $\gamma=2$ (gray) and $\gamma =1.9$ [dark (red)].
Comparison with experimental traces in Fig.\,\ref{i}(b) shows good agreement in terms of the period, the phase, the amplitude ($r_D=\rho_D \sim 1$ $\Omega$), and the slow decay with $\epsilon$.

To see if the model also catches the suppression in the regime of separated LLs, we describe the DOS by a sum of Lorentzians, broadened by $\Gamma$, and integrate Eq.\,(1) numerically. 
The result, presented in Fig.\,\ref{fig4}(b) for $\oc \tau_q=10$, shows qualitative agreement with experimental data [cf. Fig.\,\ref{sll}(b)]. 
We therefore conclude that the proposed model manages to capture the main experimental features in both regimes. It is, however, very desirable to develop a systematic theory which would also examine the role of nonequilibrium function effects.

It is interesting to compare MIRO and HIRO from experimental and theoretical perspectives.
Experimentally, the two phenomena exhibit a great deal of similarity, exploiting periodicity of the electronic spectrum of very high LLs and relying on inter-LL transitions.
In addition, in the regime of separated LLs, both microwave and dc excitations can induce strong suppression in resistance.
On the other hand, theoretically, MIRO are currently best understood by assuming only smooth disorder, while HIRO require sharp scatterers to account for $2\rc$ transitions.
It would be interesting to see if a theory treating both phenomena on the same footing becomes available.

In summary, we have studied magnetotransport in high-mobility 2DESs driven by dc electric fields in the regimes of overlapped and separated LLs.
A systematic study of Hall field-induced resistance oscillations revealed resonant condition where maxima in the differential resistance occur close to integer values of $\epsilon$, which qualitatively differs from previous reports.
In the regime of separated LLs and relatively weak dc excitation we have observed dramatic suppression of resistance which hints at the possibility of dc-induced zero-resistance states.
Both findings suggest phenomenological similarity of magnetotransport in dc-driven and microwave-driven 2DESs, prompting the question of whether these two effects are closely related.
A simple model relying on large momentum transfer is able to account for the main experimental observations but detailed theoretical study is needed to identify other possible mechanisms. It is particularly interesting to see if a theory which can treat both microwave and dc excitations within a single framework becomes available.
Finally, since even small dc currents strongly modify transport properties, dc-induced effects warrant serious consideration in relation to the ongoing search for the critical current theoretically ascribed to microwave-induced ZRS.

Recently we learned about a report on dc current-induced suppression of resistance\citep{zhang:up} which was attributed to the effect of a nonequilibrium distribution function, as opposed to the impurity scattering processes advocated in the present work. 

We thank L. Glazman, A. Kamenev, E. Kolomeitsev, B. Shklovskii, and I. Dmitriev for discussions. This work was supported by NSF Grant No. DMR-0548014.


\begin{thebibliography}{23}
\bibitem[{not({\natexlab{a}})}]{note2}
\bibinfo{note}{It is easy to show that, at least for $\epsilon \gtrsim 1$,
  oscilations in $R$ and $r$ are phase shifted by a quarter period.}
\bibitem[{not({\natexlab{b}})}]{note3}
\bibinfo{note}{This value consistently describes our data throughout the
  $B$ range studied, from 0.3 to 2 kG.}
\end{thebibliography}



\end{document}